\documentclass[12pt]{extarticle}
\usepackage{nopageno}
\usepackage{amssymb}
\usepackage{amsmath}
\usepackage[hyphens,spaces,obeyspaces]{url}
\usepackage{hyperref}
\usepackage{dsfont}
\usepackage[T1]{fontenc}
\begin{document}

\title{Is a Quantum Gravity Era Necessary?}

\author{Bogdan Veklych}

\maketitle

\textsuperscript{Author affiliation: University of Wisconsin-Madison; veklych@wisc.edu, veklych@alum.mit.edu;} 

\textsuperscript{orcid.org/0000-0003-1897-7998}

\begin{abstract}
We present the first published framework of the entirety of cosmological history which is thoroughly classical (without any quantum-gravitational era or singularities) and which passes all the known extensive consistency checks on such a model, and discuss some of its possible cosmological implications, such as its ability to account for the matter-antimatter asymmetry, dark flow, and the Hubble tension, albeit at the cost of further assumptions.
\end{abstract} 

\textsuperscript{Keywords: cosmological evolution, past-eternal cosmology, dark matter, Hubble tension}

\section{Introduction}

As is well-known, the current established cosmological knowledge of the history of the Universe extends only 13.8 billion years into the past to a hot dense rapidly expanding phase, nicknamed the Big Bang, which in turn is widely conjectured to have been produced by the decay of an inflationary field - but further than that it is not known with any certainty what preceded it or how it came about. While it is easy to come up with speculative quantum-gravitational ideas for the ultimate past of the Universe (such as e.g. origin of Universes by budding from "quantum foam" in an eternal empty progenitor Universe, or, as another example, an empty 2+1-dimensional compactified Milne past-eternal contracting phase, stable due to rigidity of 2+1-dimensional vacuum, which then changes dimensionality at the bounce via some hypothetical quantum-gravitational mechanism), apparently there has never been an example in the literature of a thoroughly classical consistent cosmological model of the full prehistory of the cosmic inflation, with no singularities or speculative quantum-gravitational epochs or other handwaved unclear "gaps in history". This is however not surprising as there are numerous very tight consistency constraints and no-go results on such a model [1][2][3][4][5][6], and the purpose of the present article is to fill this theoretical gap by introducing a framework with such desirable features that satisfies all these consistency constraints.

\section{The General Framework and a Particular Toy Model Illustration} 

The basic description of the model is as follows. In the pre-inflationary "burning fuse era", the Universe has two spatial dimensions compactified (i.e. it is shaped like an "infinite corridor" with flat torus cross-sections), with two waves approaching each other along the third dimension with the speed of light from past-eternity, while the Universe is static, flat, and empty in the shrinking middle part between these two waves. When these primordial waves finally collide and react, part of their contents forms proto-inflationary scalar field(s), and another part is matter with nonzero intrinsic angular momentum (like e.g. ordinary photons with spin). The Universe may start collapsing, but due to the matter it bounces back - here we, crucially, assume validity of Einstein-Cartan theory - after which the true inflation starts. (It is very plausible that this setup can be slightly altered to have a pure Einstein-Cartan bounce with no inflation after the wave collision, but for simplicity we won't consider this here.)

We shall now give a particular proof-of-concept toy illustration of this model, with some additional simplifying assumptions, and then discuss the consistency of our past-eternal cosmology. Note that a long chain of highly ad hoc assumptions, akin to an eternal Rube Goldberg machine, is fundamentally inevitable in any realization of this basic idea (and this is the trade-off for really avoiding quantum gravity); see the end of the next section for why this is completely benign. Our starting point is the effect that a perfect fluid with spin together with a scalar field have on the metric in Einstein-Cartan theory (see the equation (77) in [7]):

\begin{equation}\begin{aligned}G_{kl}=\kappa\left(\rho-\frac{1}{8}\kappa s_{bd}s^{bd}\right)u_ku_l+\kappa\left(p-\frac{1}{8}\kappa s_{bd}s^{bd}\right)(g_{kl}+u_ku_l)\\ [+\kappa\Theta_{kl}]-\frac{1}{2}\kappa(g^{bd}+u^bu^d)\nabla_d(s_{bk}u_l+s_{bl}u_k) \end{aligned}\end{equation}

\noindent where $G_{kl}$ is the Einstein tensor (in the ordinary sense, constructed from the Levi-Civita connection, likewise for the covariant derivative), $\kappa=8\pi G/c^4$ (henceforth we set $\kappa=c=1$), $g_{kl}$ is the metric tensor, $\rho, p, u^k, s_{kl}$ are the density, pressure, four-velocity, and spin density of the fluid, correspondingly, and $\Theta_{kl}$ is the (canonical and symmetric) stress-energy tensor of a scalar field $\phi$ (which can be incorporated into $\rho$ and $p$, hence the brackets). From (1) we can derive the effect on the metric of spinning null dust by a simple limit procedure, as follows. Let the spin and velocity point along the $x$-direction, i.e. let $s_{kl}=0$ for $k+l\neq 5$ and $u^2=u^3=0$ (note that the expression $u^bu^d\nabla_d(s_{bk}u_l+s_{bl}u_k)$ now vanishes), and take the limit as $u^0\rightarrow \infty$ while keeping $(u^0)^2\rho, (u^0)^2 p, u^0 s_{kl}$, and $u^1/u^0$ constant. In the end, we obtain 

\begin{equation}\begin{aligned}G_{kl}=\Theta_{kl}+\left(\epsilon-\frac{1}{4} s_{bd}s^{bd}\right)n_kn_l-\frac{1}{2}g^{bd}\nabla_d(s_{bk}n_l+s_{bl}n_k) \end{aligned}\end{equation}

\noindent where $n^k$ is a null vector and $\epsilon (n^0)^2$ is the energy density of the null dust. Further, in the case of a simple plane wave characterized by $n^0=1, n^1=-1, s_{23}=-s_{32}=s(x+t), \epsilon = \epsilon(x+t), g_{kl}=\operatorname{diag}(-1, 1, h(x+t)^2, h(x+t)^2), \phi=\phi(x+t)$, (2) reduces to \begin{equation}-2\frac{h''}{h}=(\phi')^2+\epsilon-\frac{s^2}{2h^4}\end{equation}

\noindent With an increasing numerical density of identical particles, $s^2$ grows faster than $\epsilon$, which means that for concentrated enough matter the right hand side of (3) can become negative, making the space that these particles reach expand (or "bounce") transversally. For maximum simplicity, let's hold the matter concentration at exactly "the brink of bounce", furthermore taking $\epsilon$ to be negligibly small, reducing the metric to flat ($h=\operatorname{const}$, say $h=1$) and (3) to $(\phi')^2=\frac{1}{2}s^2$, and similarly for the left wave. Let's also assume that once this right wave collides with the symmetric left wave at $t=0$, the intermediate interaction region $t>0, |x|<t$ will be homogeneous (with all quantities dependent at most on $t$) and flat (but anisotropic - having a preferred direction, in which the spin is pointing), which, by (1), is achievable if \begin{equation}\frac{1}{4}s(t)^2=\frac{1}{2}(\phi(t)')^2
\end{equation} 
\noindent (as $\rho=p=\frac{1}{2}(\phi(t)')^2$ for the scalar field). To this end, let's make the following further toy model assumptions: the spin density in the waves is constant, and so is the kinetic energy of the scalar field (i.e. the field is linearly changing with $t\pm x$); the spinning particles immediately stop and become spinning dust when reaching the middle region (for example, due to a reaction that amounts to an inelastic collision with some spinless particles traveling in the opposite direction). Now observe that 1) under these conditions spin becomes twice more concentrated in the middle region than in the waves - a spinning particle in a wave which is at a distance $2\delta$ away from the interaction front will move $\delta$ more before being stopped - thus $\frac{1}{4}s(t)^2$ in the middle region is twice larger than $\frac{1}{2}s^2$ in the waves; 2) if $\phi(x+t)=f_1(x+t)$ in the right wave and $\phi(t)=f_2(t)$ in the middle region, from the fact that they match at $t=x$ we conclude that $f_2(t)=f_1(2t)$, thus $\frac{1}{2}(\phi(t)')^2$ in the middle region is twice larger than $(\phi')^2$ in the waves. These two observations imply that (4) is satisfied in the intermediate region if (3) was satisfied in the waves, and so our assumptions are consistent. (As a reminder, we assumed, for simplicity, that up to this point the metric was flat, which simplified the equations (1)-(3) immensely.)

Now, to enable the beginning of inflation, we assume (yet further) that the scalar field has a stepfunction potential, so that it is only effectively massless as long as it is below a certain value. (A more natural, but less computationally convenient, option is to take two scalar fields $\psi_{left}$ and $\psi_{right}$ with a combined polynomial or other smooth potential that includes a factor of $\psi_{left}^2\psi_{right}^2$, so that either one is effectively massless when the other one is zero, and have $\psi_{left}$ be zero to the right of the left wave and $\psi_{right}$ be zero to the left of the right wave.) Specifically, assume that when it hits that value at some moment $t_0$ (which can be as large as needed) all of its kinetic energy becomes potential energy (and we also assume that in the waves the growth of $\phi$ is truncated at infinitesimally less than that value). From now on we're only interested in the region causally uninfluenced by anything outside of the homogeneous middle part at the moment of this transition. By definition this region is shrinking inward with the speed of light on its right and left edges, and internally it is subject to FLRW-like equations (see [8]; in the following $a(t)=g_{11}=g_{22}=g_{33}$ is the scale factor, depending only on time, $a(t_0)=1$, $a'(t_0)=0$, $V$ is the height of the potential of the scalar field, and $s_0^2$ is constant)

\begin{equation}
3\frac{(a')^2}{a^2}+\frac{s_0^2}{4a^6}=V=2\frac{a''}{a}+\frac{(a')^2}{a^2}-\frac{s_0^2}{4a^6}
\end{equation}

From the first equality in (5) we get $(a'/a)^2\leq V/3$, substituting this into the second equality we get $a''/a\geq V/3$; in terms of $A(t)=\ln a(t)$ this reads $(A')^2+A''\geq V/3$, thus $A'$ increases from zero at first and cannot decrease if it is less than $\sqrt{V/3}$, so $A$ is bounded below by a positively-sloped line, and therefore $a$ (eventually) grows exponentially. From this it also follows that, if $t_0$ is large enough, this region will expand exponentially in the $x$-direction as well (that is, with its linearly inward-moving left and right boundaries never overtaking its exponential expansion in the middle) - the (toy-model) inflation has started.

\section{Consistency}

In this section we show how this general framework (with any secondary details) circumvents the numerous tight consistency constraints mentioned before.

Penrose-Hawking singularity theorems [1] do not apply to Einstein-Cartan theory. 

Compactification of all but one spatial dimensions automatically stabilizes and homogenizes plane waves. For example, suppressing one spatial dimension for convenience (and limiting ourselves only to special relativity), consider an arbitrary "splash" of a massless scalar field $\psi$ in a flat 2+1 dimensional Universe with one compactified spatial dimension of (minimal incontractible loop) length $2\pi$, in the appropriate units of length. Applying the Klein-Gordon equation $\square \psi=0$ to the Fourier decomposition of $\psi$ along the periodic coordinate, $\psi(t,x,y)=\sum_{k=-\infty}^\infty \psi_k(t,x)e^{i k y}$, we get the set of equations $(\square_{t,x}-k^2)\psi_k(t,x)=0, k \in \mathds{Z}$. In particular, taking $k=0$, we see that $\psi_0(t,x)=\frac{1}{2\pi}\int_0^{2\pi}\psi(t,x,y)dy$ satisfies the one-dimensional wave equation for speed 1 and (after possibly splitting into left- and right- moving parts) will be indefinitely stably propagating.

There is no quantum dispersion [2] because any relevant wavefunction is relativistically "frozen" by motion with the speed of light. And even Heisenberg uncertainty inequality does not apply in the case of a periodic coordinate: for example, in the textbook "particle in a ring" scenario the particle can have zero angular momentum, even though its angular position is only finitely imprecise (having the maximum possible imprecision, $2\pi$).

This model doesn't have $H>0$ [3] (nor $H<0$ [4] with its environment of unlimited blueshift, where any fluctuation - vanishing in the limit of past-infinity - becomes deadly magnified). In fact, in this model any geodesically moving observer was always, before a certain moment, simply in static flat empty space. (Note, importantly, that any proposals in which the Universe has a "life cycle", e.g. originating in a black hole in an older Universe, which did likewise, "and so on", do not pass [3].)

This model also gives an example of how it is possible for "the Universe [to be] in a near-equilibrium state for the first "half" of eternity and then ... [to exit] this equilibrium" ([5], p.29); indeed, it can be explicitly shown that there is no entropy change (in particular, no entropy decrease) before the wave collision, as follows. Considering the effect of one of the waves, we can replace distant causally not yet reached regions, including those with the other wave, by more of empty space compactified in two dimensions; now, this Universe looks exactly the same after some time as before - if we just shift our vantage point by the distance the wave has traveled - so its entropy is also the same. Thus we conclude that there is no entropy change in the pre-collision era even under some yet-unknown possible future quantum-gravitational understanding of entropy, in the spirit of [5].

Static flat empty space is not subject to the quantum instabilities of [6] (confirmed by A. Vilenkin in personal correspondence) due to fundamental problems with assigning consistent decay-governing wavefunctions to it. And, finally, we also need to make the very reasonable conjecture that static flat empty space does not spontaneously produce matter. (It would be quite a challenge to disprove it - such a disproof would yield the simplest \emph{perpetuum mobile} design of all, just a large empty container, inside of which matter is spontaneously generated and collected!) Another way to put this point is that quantum fluctuations in static flat vacuum do not increase entropy.

Note that this model is extremely conservative (in particular, the characteristic peak density reached during an Einstein-Cartan bounce, called the Cartan density, is about forty orders of magnitude less than the Planck density, so quantum gravity is not relevant during the wave collision), corresponding precisely to the classical picture argued for by natural philosophers like Aristotle and Giordano Bruno: time, space, motion, and matter have existed forever, with matter always coming only from the matter that existed in the preceding moments. (As to its fine-tuning, it can be assumed, for example, that our Universe is a part of a larger Multiverse with all sorts of physical parameters in different places - a common suggestion, also applicable to this model; this "anthropic principle" can cover any remaining loose ends in the model that might seem strange or unlikely but are not outright demonstrably excluded - which is why the focus of the above was pure consistency. Any particular realization of the idea above will be something that can be described impressionistically as messy, ad hoc, and convoluted - but arguably the same can be said about the Standard Model of particle physics, for example, so this is not an argument for its falsity.)

\section{Possible cosmological implications?}

If we additionally assume that inflation only existed briefly (rather than as eternal chaotic inflation, which would dilute away any traces of its prehistory), many intriguing possibilities open up if this framework is correct - but, as always with this "Rube Goldberg cosmology", at the cost of further convoluted assumptions: 

1) Let's assume that the lightest neutrino mass eigenstate is, surprisingly enough, exactly zero, which in turn enables us to make another assumption - that the lightest antineutrino was present in the primordial waves, in greater abundance than the corresponding neutrino. Then the matter-antimatter imbalance is trivially accounted for under our framework - there was simply always more matter (with positive $B-L$, preserved by sphalerons) than antimatter, both before and after the collision of the primordial waves!

2) Let's also assume that dark matter, likewise, had its own lightspeed precursor(s) in the primordial waves, which reacted upon collision and formed (nonzero-mass) dark matter - and that neither this precursor nor dark matter itself interacts with luminous matter other than gravitationally. Then there is no reason to assume that the inertial frame in which the average momentum of luminous matter post-collision was zero was the same as the corresponding frame for dark matter. In other words, we can assume that (massive) dark and luminous matter originally were moving with respect to each other in the direction of the uncompactified dimension, and could have kept some of this relative motion even at the time when the cosmic microwave background radiation was emitted. Later, by mutual local gravitational tugs, they would naturally finally "stick together" and equalize their average velocities - which would mean that the visible matter would be now slightly moving on average with respect to the inertial system defined by the cosmic microwave background radiation (which is the system reflecting its motion before the slight residual "final adjustment" by dark matter). But this is indeed precisely what we observe, as the "dark flow" phenomenon. In other words, our framework provides an explanation for the mysterious dark flow - and if this explanation is correct, then, in turn, the dark flow gives us the direction of the uncompactified dimension.

3) Finally, let's furthermore assume that dark matter is "fuzzy" (which has been recently confirmed [9]) but not scalar (that is, its particles are extremely light and have spin 1 or 2), and also (similarly to the speed imbalance assumption in 2) above) that the spins of the dark matter lightspeed precursors weren't precisely balanced to zero average in the colliding primordial waves. Then dark matter now has net nonzero intrinsic angular momentum pointing in the direction of the uncompactified dimension, inherited from its primordial wave lightspeed progenitors. If so, the Universe is homogeneous but not isotropic, having a preferred direction (in which the intrinsic angular momentum of fuzzy dark matter is pointing), and is described by [8] instead of a FLRW metric - which can explain the Hubble tension neatly [10], and this explanation is even potentially experimentally checkable, via the observational consequences of the corresponding spacetime torsion!

That is, if dark matter is fuzzy and has intrinsic angular momentum (pointing in the direction of the dark flow or opposite to it, by 2), under our chains of assumptions), then the standard cosmological Lambda-CDM model can be augmented, with the relevant FLRW-like equations becoming (1) in [8], and the torsion produced by the corresponding nonzero value of $\mu=K_0^2$ (in the notations of [8]) is a prediction that can (and should!) be tested experimentally [11]. 

In fact, as the dark flow demonstrates, there might indeed be a preferred direction in the Universe, and if the assumption of its isotropy is dropped, then the above one-parameter extension of the Lambda-CDM model is by far the simplest possible explanation of the Hubble tension even without the corroborating background provided by the general classical framework proposed in this article, and therefore it is a reasonable candidate for being given an observational test. (And there might also be other independent corroborating evidence explained by this model, such as hints of a macroscopic compactified dimension [12].)\\

\noindent \textbf{References}\\

[1] S. W. Hawking, "Singularities in the Universe", \emph{Phys. Rev. Lett. 17}, 444

[2] A. Aguirre, J. Kehayias, "Quantum instability of the emergent universe", \emph{Physical Review D, 88(10)}, arxiv.org/abs/1306.3232

[3] A. Borde, A. H. Guth, A. Vilenkin, "Inflationary spacetimes are not past-complete", \emph{Phys. Rev. Lett. 90} (2003) 151301, abs/gr-qc/0110012; 

W. H. Kinney, N. K. Stein, "Cyclic Cosmology and Geodesic Completeness", \emph{JCAP} (accepted for publication), abs/2110.15380 

[4] T. Banks, W. Fischler, "Black Crunch", abs/hep-th/0212113

[5] A. C. Wall, "The Generalized Second Law implies a Quantum Singularity Theorem", \emph{Class. Quantum Grav. 30}, 165003 (2013), abs/1010.5513 

[6]  A. T. Mithani, A. Vilenkin, "Collapse of simple harmonic universe", abs/1110.4096; 

A. T. Mithani, A. Vilenkin, "Did the universe have a beginning?", abs/1204.4658

[7] K. Pasmatsiou, C. G. Tsagas, J. D. Barrow, "Kinematics of Einstein-Cartan universes", \emph{Phys. Rev. D, 95(10)}:104007 (2017), abs/1611.07878

[8] W. Kopczynski, "An anisotropic universe with torsion", \emph{Phys. Lett. A, 43(1)}, 63-64 (1973) 

[9] A. Amruth et al., "Anomalies in Gravitational-Lensed Images Revealing Einstein Rings Modulated by Wavelike Dark Matter", \emph{Nat. Astron. (2023)}, abs/2304.09895

[10] F. Izaurieta, S. Lepe, O. Valdivia, "The Spin Tensor of Dark Matter and the Hubble Parameter Tension", abs/2004.13163; 

T. Liu et al., "Revisiting Friedmann-like cosmology with torsion: newest constraints from high-redshift observations", \emph{JCAP} (accepted for publication), abs/2304.06425

[11] K. Bolejko et al., "Cosmological signatures of torsion and how to distinguish torsion from the dark sector", \emph{Phys. Rev. D, 101} (2020) 104046, abs/2003.06528

[12] Y. Akrami et al., "The Search for the Topology of the Universe Has Just Begun", abs/2210.11426

\end{document}